\documentstyle[12pt]{article}

\newcommand{\C}{\bf{C}}
\newcommand{\R}{\bf{R}}

\newcommand{\bz}{\bar{z}}

\newcommand{\bw}{\bar{w}}
\newcommand{\hN}{\hat{N}}

\renewcommand{\H}{{\cal H}}
\newcommand{\bea}{\begin{eqnarray}}
\newcommand{\ena}{\end{eqnarray}}
\newcommand{\EQ}{\begin{equation}}
\newcommand{\EN}{\end{equation}}
\newcommand{\vs}[1]{\vspace{#1 mm}}

\newcommand{\pa}{\partial}
\newcommand{\nn}{\nonumber\\}
\newcommand{\lan}{\langle}
\newcommand{\ran}{\rangle}

\makeatletter
 
 \@addtoreset{equation}{section}
\makeatother

\begin{document}


\topmargin 0pt
\oddsidemargin 0mm
\renewcommand{\thefootnote}{\fnsymbol{footnote}}

\begin{titlepage}

\setcounter{page}{0}
\begin{flushright}
KEK-TH 720\\
hep-th/0010119\\
\end{flushright}

\vs{20}
\begin{center}
{\Large\bf  Dp-D(p+4)
in Noncommutative Yang-Mills}\\
\vs{20}
{\large
Furuuchi \ Kazuyuki}

\vs{10}
{\em Laboratory for Particle and Nuclear Physics,}\vs{1}\\
{\em High Energy Accelerator Research Organization (KEK), }\vs{1}\\
{\em Tsukuba, Ibaraki 305-0801, Japan} \vs{2}\\
{\em Fax : (81)298-64-5755} \vs{1}\\
{\em E-mail: furuuchi@post.kek.jp}
\end{center}
\vs{10}
\renewcommand{\thefootnote}{\fnsymbol{footnote}}
\setcounter{footnote}{0}
\centerline{{\large\bf Abstract }}\vs{5}%
\noindent
An anti-self-dual instanton solution in Yang-Mills theory
on noncommutative ${\R}^4$ with an anti-self-dual
noncommutative parameter is constructed.
The solution is constructed 
by the ADHM construction
and it can be treated in the
framework of the IIB matrix model.
In the IIB matrix model, this solution is interpreted
as a system of a Dp-brane and D(p+4)-branes, with
the Dp-brane dissolved in the worldvolume of the 
D(p+4)-branes.
The solution has a parameter that characterises
the size of the instanton.
The zero of this parameter
corresponds to the singularity of the moduli space.
At this point, 
the solution is continuously connected to 
another solution which 
can be interpreted as a system of 
a Dp-brane and D(p+4)-branes, with
the Dp-brane separated from the D(p+4)-branes.
It is shown that even when the parameter of 
the solution comes to the
singularity of the moduli space,
the gauge field itself is non-singular.
A class of multi-instanton solutions 
is also constructed.

\end{titlepage}
\newpage

\renewcommand{\thefootnote}{\arabic{footnote}}
\setcounter{footnote}{0}
\section{Introduction}
After the discovery of Dirichlet Branes (D-branes)
\cite{Dbrane} 
the nonperturbative analysis of string theory
has achieved much progress.
One of the key features here is that
the low energy effective field theory on the worldvolume 
of Dp-branes
describes the configuration of the D-branes in a target space,
and vice-versa
\cite{bound}. 

The Dp-D(p+4) system has attracted much interest
for a number of reasons.
The ground states of the Dp-D(p+4) 
system preserve one-fourth of the supersymmetry 
in superstring theory.
One of the interesting features of 
this system is that
it has descriptions from two different viewpoints.
The low energy effective theory on the worldvolume 
of the D(p+4)-branes
is a supersymmetric Yang-Mills theory, and
when the Dp-branes are within the worldvolume of the D(p+4)-branes
the Dp-branes are described as instantons on ${\R}^4$ transverse
to the Dp-branes and within the D(p+4)-brane.
On the other hand,
the low energy effective field theory on the worldvolume of
the Dp-branes is in the Higgs branch,
and the moduli space 
coincides with the moduli space of 
instantons on ${\R}^4$
\cite{SI}\cite{pinp4}.\footnote{
Here we are considering the classical moduli space.
For the discussions on possible quantum corrections,
see \cite{singCFT}.}
The moduli space of instantons has so-called
small instanton singularities.
These correspond to the instantons shrinking to
zero-size, and the 
low energy Yang-Mills description on the D(p+4)-branes
may break down.
In the Dp-brane worldvolume theory, the
Higgs branch meets
the Coulomb branch
at these small instanton singularities. 
The Coulomb branch describes the separation
of the Dp-branes from the D(p+4)-branes
in the direction transverse to the D(p+4)-branes.

The constant NS-NS B-field background in the
worldvolume of D(p+4)-branes gives interesting effects
to this Dp-D(p+4) system.
Under the constant NS-NS B-field background
the coordinates on the D(p+4)-branes become
noncommutative.
On the other hand,
when the B-field in the ${\R}^4$ has a 
non-zero self-dual part
(in our convention the 
Dp-branes are described as anti-self-dual instantons),
the field theory on Dp-branes acquires
a Fayet Iliopoulos term
\cite{ABS}.
Then, the
small instanton singularities in the moduli space
are resolved 
\cite{iNakaj}, and the Coulomb branch disappears
from the field theory on the worldvolume of the Dp-branes.
This means that the Dp-branes are confined within 
the D(p+4)-branes.
Since the coordinates on the D(p+4)-branes become 
noncommutative,
the Dp-branes should be described as instantons on 
{\it noncommutative} ${\R}^4$
in the worldvolume theory on the D(p+4)-branes.
This expectation is confirmed by the beautiful results
in \cite{NS};
the moduli space of the field theory on the worldvolume
of the Dp-branes under this background
coincides with the moduli space
of instantons on noncommutative ${\R}^4$.

Recently,
a classical solution of 
the Yang-Mills theory on noncommutative ${\R}^4$
(noncommutative Yang-Mills)
was found in \cite{AGMS}.
It is a self-dual gauge field configuration
on noncommutative ${\R}^4$ with a self-dual
noncommutative parameter.
This solution can be realized in 
the M(atrix) model \cite{BFSS}\cite{CDS} and
the IIB matrix model \cite{IKKT}\cite{SFTIIB},
and then it coincides with a solution which has been
studied before \cite{IIBD}.
In these matrix models, 
this solution is interpreted as  
a system of Dp-branes and D(p+4)-branes separated from
each other.

Motivated by \cite{AGMS},
in this article
we construct a new \footnote{
The existence of such solution itself
is a rather straightforward
consequence of the ADHM construction.
However, the explicit expression of 
the small instanton limit may be non-trivial.
} $U(2)$
anti-self-dual instanton solution
on noncommutative ${\R}^4$ with
an anti-self-dual noncommutative parameter.\footnote{%
The case of anti-self-dual gauge configuration on
${\R}^4$ with
self-dual noncommutativity has
already been studied in some detail
(see \cite{NS},\cite{mine}-\cite{NLec}).
}
The construction of the
anti-self-dual gauge configuration
on ${\R}^4$ with 
anti-self-dual noncommutativity is technically
the same as the
construction of a self-dual gauge configuration
on ${\R}^4$ with 
self-dual noncommutativity.
The solution is constructed by the
ADHM construction, and
can be treated in 
the framework of the IIB matrix model.
In the IIB matrix model, this solution
can be interpreted as a system of a Dp-brane and 
D(p+4)-branes, with
the Dp-brane dissolved in the
worldvolume of the D(p+4)-branes, and
with constant anti-self-dual NS-NS B-field background
in the worldvolume of the D(p+4)-branes.
In this case, the
moduli space of instantons has a small instanton
singularity.
When the moduli parameter of the
solution comes
at this singularity, the solution is
continuously
connected to the solution discussed
in \cite{AGMS}\cite{IIBD}.
It is explicitly shown that
even though the parameter of the 
solution comes to the 
singularity of the moduli space,
the corresponding gauge configuration is non-singular.
Thus, the noncommutative Yang-Mills can describe
the separation of a Dp-brane off the D(p+4)-branes,
with non-singular variables.
This is quite remarkable compared to the 
commutative case,
because instantons
on commutative space
can describe only Dp-branes within the
worldvolume of D(p+4)-branes,
since 
the field strength
becomes singular at the small instanton singularity.

We also make some comments on 
the instanton position parameter.
When the size of the instanton becomes minimal,
the instanton position parameter can be
identified, in the IIB matrix model, 
with a position of the  
D(-1)-brane in the direction parallel to the D3-brane
worldvolume.

A class of multi-instanton solution is also constructed.
The origin of the small instanton singularities can be
observed directly from the field configuration.

\newpage
\section{Yang-Mills Theory on 
Noncommutative ${\bf R}^4$  }\label{sg}
In this section we briefly review
Yang-Mills theories  on noncommutative ${\R}^4$
and their appearance in the IIB matrix model
with certain backgrounds.
\subsection*{Gauge Fields on Noncommutative ${\R}^4$}
The coordinates 
$x^{\mu} \, \,  (\mu  = 1 ,\cdots , 4)$ of 
the noncommutative
${\R}^4$ obey the
following
commutation relations:
\bea
 \label{noncomx}
[ \hat{x}^{\mu} ,  \hat{x}^{\nu}] = i \theta^{\mu\nu} , 
\ena
where the noncommutative parameter
$\theta^{\mu\nu}$ is a real 
constant matrix.
By $SO(4)$ rotation in ${\R}^4$ one can set the 
components of the matrix $\theta^{\mu\nu}$ to
zero, except $\theta^{12} = -\theta^{21}$ and 
$\theta^{34} = - \theta^{43}$.
We introduce the complex 
coordinates by 
\bea
z_1 = \hat{x}^2 + i \hat{x}^1  , \quad 
z_2 = \hat{x}^4 + i \hat{x}^3 \,  .
\ena
Their commutation relations become
\bea
 \label{noncom}
& &[z_1 , \bar{z}_1] = \zeta_1, \quad 
[z_2 , \bar{z}_2] = \zeta_2, \nn
& &[z_1 , z_2] = [z_1 , \bz_2] = 0,
\ena
where
$\zeta_1 = - 2\theta^{12}$ and $\zeta_2 = - 2\theta^{34}$. 
In this article we study the case where
$\theta^{\mu\nu}$ is anti-self-dual, i.e.
$\theta^{12} + \theta^{34} =0$.
This means $\zeta_1 = -\zeta_2 $. 
Further, we set $\zeta_1 > 0$.
We then define
\bea
a_1 \equiv \sqrt{\frac{1}{\zeta_1}} z_1, \quad & &
a_1^{\dagger} \equiv \sqrt{\frac{1}{\zeta_1}} \bz_1 \, , \\
a_2 \equiv \sqrt{\frac{1}{\zeta_1}} \bz_2, \quad & &
a_2^{\dagger} \equiv \sqrt{\frac{1}{\zeta_1}} z_2 \, .
\ena
We realize
$a^{\dagger}$ and $a$ 
as creation and annihilation operators
acting in a Fock space $\H$ spanned
by the basis
$\left| n_1 , n_2    \right\ran$:
\bea
 \label{basis}
a_1^{\dagger} \left| n_1  , n_2    \right\ran
&=& \sqrt{n_1 +1} \left| n_1 + 1 , n_2    \right\ran , \quad
a_1 \left| n_1  , n_2    \right\ran
= \sqrt{n_1 } \left| n_1 -1 , n_2    \right\ran , \nn
a_2^{\dagger} \left| n_1  , n_2    \right\ran
&=& \sqrt{n_2 +1} \left| n_1 , n_2 +1    \right\ran , \quad
a_2 \left| n_1  , n_2    \right\ran
= \sqrt{n_2} \left| n_1 , n_2 -1    \right\ran .
\ena
The commutation relation (\ref{noncomx})
has automorphismes of the form
$ \hat{x}^{\mu} \mapsto \hat{x}^{\mu} + y^{\mu}$
(translation),
where $y^{\mu}$ is a commuting real number.
We denote the Lie algebra of this group
by ${\bf \underline{g}}$.
These automorphismes are
generated by the unitary operator $T_y$
\bea
 \label{tra}
T_y \equiv
\exp
[y^{\mu} \hat{\pa}_{\mu}] ,
\ena
where we have introduced a
{\bf derivative operator}
$\hat{\pa}_{\mu}$ by
\bea
 \label{deri}
\hat{\pa}_{\mu} \equiv i B_{\mu\nu} \hat{x}^{\nu}.
\ena
Here, $B_{\mu\nu}$ is an
inverse matrix of $\theta^{\mu\nu}$.
The derivative operator 
$\hat{\pa}_{\mu}$ satisfies the following
commutation relations:
\bea
 \label{comdel}
[\hat{\pa}_{\mu}, \hat{x}^{\nu}] = \delta_{\mu}^{\nu}, \quad
[\hat{\pa}_{\mu}, \hat{\pa}_{\nu}]
= i B_{\mu\nu}.
\ena
From (\ref{comdel}) we obtain 
\bea
T_y\, \hat{x}^{\mu}\, T_y^{\dagger} 
=\,  \hat{x}^{\mu} + y^{\mu}.
\ena
For any operator $\hat{O}$
we define {\bf derivative of operator} $\hat{O}$ by 
the action of \underline{\bf g}:
\bea
\label{del}
\pa_{\mu} \hat{O} \equiv
\lim_{\delta y^{\mu} \rightarrow 0}
\frac{1}{\delta y^{\mu} } 
\left(
T_{\delta y^{\mu}}
\hat{O} T_{\delta y^{\mu}}^{\dagger} - \hat{O}
\right)=
[\hat{\pa}_{\mu}, \hat{O} ].
\ena
The action of the exterior derivative 
$d$ to the operator $\hat{O}$ is 
defined as
\bea
d\hat{O} \equiv (\pa_{\mu} \hat{O} )\,dx^{\mu}. 
\ena
Here, $dx^{\mu}$'s
are defined in the usual way, i.e.
they commute with
$\hat{x}^\mu$ and anti-commute among themselves:
$dx^{\mu}dx^{\nu} = -dx^{\nu}dx^{\mu}$.
The covariant derivative $D$ is
written as
\EQ
D = d+A.
\EN
Here, $A= A_\mu dx^\mu$ 
is a $U(n)$ gauge field.
$A_\mu$ is an $n \times n $ anti-Hermite
operator-valued matrix.
The field strength of $A$ is given by
\bea
F \equiv D^2 = dA + A^2 
\equiv \frac{1}{2} F_{\mu\nu} dx^\mu dx^\nu.
\ena
We consider the following 
Yang-Mills action
\bea
 \label{SO}
S = \frac{1}{4g^2}
\left(
\pi \zeta_1
\right)^2
\mbox{Tr}_{{\cal H}}\mbox{tr}_{U(n)}\,
F_{\mu\nu} F^{\mu\nu} .
\ena
The action (\ref{SO})
is invariant under the following
$U(n)$ gauge transformation:
\bea
 \label{gauget}
A \rightarrow 
UdU^{\dagger} +
UAU^{\dagger}  .
\ena
Here, $U$ is a unitary operator:
\bea
UU^{\dagger} = U^{\dagger}U =
\mbox{Id}_{\H} \otimes \mbox{Id}_{n} \, ,
\ena
where $\mbox{Id}_{\H}$ is the identity operator
acting in $\H$ and $\mbox{Id}_{n}$ is 
the $n \times n$ identity matrix.
We will also simply write this kind of
identity operators
as ``1'' , if this is not confusing. 
The gauge field $A$ is called
{\bf anti-self-dual}
if its field strength obeys the
following equation:
\bea
 F^+ \equiv
 \frac{1}{2} 
 (F + * F) = 0 ,
\ena
where
$*$ is the Hodge star.\footnote{In this note
we only consider the case
where the metric 
on ${\R}^4$ is flat:
$g_{\mu\nu} = \delta_{\mu\nu}$.}
Anti-self-dual gauge fields minimize
the Yang-Mills action (\ref{SO}).
An instanton is an anti-self-dual
gauge field with finite
Yang-Mills action (\ref{SO}).

One can consider a one-to-one map
from operators to
ordinary c-number functions on ${\R}^4$.
Under this map,
noncommutative operator multiplication is
mapped to the so-called star product.
The map from operators to 
ordinary functions depends on
an operator ordering prescription.
Here, we choose the Weyl ordering.

Let us consider Weyl ordered operator
of the form
\bea
 \label{Nop}
\hat{f}(\hat{x}) = 
\int \frac{d^4k}{(2\pi)^4} \, 
\tilde{f}_W (k) \,   e^{ik\hat{x} }  \, ,
\ena
where $k\hat{x} \equiv k_{\mu}\hat{x}^{\mu}$.
For the 
operator-valued 
function 
(\ref{Nop}),
the corresponding {\bf Weyl symbol} 
is defined by
\bea
 \label{symb}
f_W (x) =
\int \frac{d^4k}{(2\pi)^4} \,
\tilde{f}_W (k) \, e^{ikx}\, , 
\ena
where $x^{\mu}$'s
are commuting coordinates of ${\R}^4$. 
We define $\Omega_W$ as a
map from operators to corresponding Weyl symbols:
\bea
 \label{WeylT}
\Omega_W(\hat{f}(\hat{x}) ) = 
f_W (x).
\ena
One can show the relation
$\mbox{Tr}_{\cal H}
\bigl\{
 \exp \, ( ik\hat{x} )  \bigr\}
= \left( \pi \zeta_1 \right)^2
   \delta^{(4)}(k)$. %
We then obtain
\bea
 \label{intr}
\left( \pi \zeta_1 \right)^2
\mbox{Tr}_{\cal H}\, \hat{f} (\hat{x})
=
\int d^4x\,  
f_W (x).
\ena
The {\bf star product} 
of functions is defined by
\bea
 \label{star}
f(x) \star g(x) \equiv
\Omega_W (\Omega_W^{-1} (f(x)) \Omega_W^{-1} (g(x))) \, .
\ena
Since 
\bea
 \label{starpro}
& & e^{ik\hat{x}} \, e^{ik'\hat{x}} =
e^{ - \frac{i}{2} \theta^{\mu\nu} 
k_\mu {k_{\nu}}'
} 
e^{ik\hat{x} + ik' \hat{x}},
\ena
the explicit form of the star product is given by
\bea
f(x) \star g(x)
=
\left.
e^{ \frac{i}{2} \theta^{\mu\nu}
\frac{\pa }{\pa x^{\mu} }
\frac{\pa}{\pa {x'}^{\nu} }
}
f(x) 
g(x')
\right|_{x' = x} \, .
\ena
From the definition (\ref{star}), the star product
is associative
\EQ
(f(x)\star g(x) \, ) \star h(x)
=
f(x)\star (g(x) \star h(x) \,).
\EN
We can rewrite (\ref{SO}) using the Weyl symbols
\bea
 \label{Ss}
S = \frac{1}{4g^2}
\int
\mbox{tr}_{U(n)}\, F  * F.
\ena
In (\ref{Ss}), multiplication of the
fields is understood  
to be the star product.
The {\bf instanton number} is
defined by 
\bea
 \label{Inum}
-\frac{1}{8\pi^2} \int \mbox{tr}_{U(n)}\,  F F \, ,
\ena
and takes an integral value.

\subsection*{Noncommutative Yang-Mills in the IIB 
Matrix Model}\label{secIIB}

It is sometimes convenient 
to treat the 
classical solution of 
noncommutative Yang-Mills theories
in the framework of the IIB matrix model.
In the IIB matrix model, noncommutative
Yang-Mills theories appear
from expansions around certain
backgrounds \cite{MLi}\cite{IKKTB}.

The IIB matrix model was
proposed as a nonperturbative formulation of
type IIB superstring theory \cite{IKKT}\cite{SFTIIB}.
It is defined by the following action:
\bea
 \label{IIB}
S = -\frac{1}{g^2}
\mbox{Tr}_{U(N)}
\left(
\frac{1}{4}[X_\mu, X_\nu][X^\mu, X^\nu]
+
\frac{1}{2}\bar{\Theta}\Gamma^\mu
[X_\mu,\Theta]
\right)\quad (\mu = 0,\cdots,9),
\ena
where 
$X_\mu$ and $\Theta$ are $N \times N$
hermitian matrices and each component
of $\Theta$ is a Majorana-Weyl spinor.
The action (\ref{IIB}) 
has the following $U(N)$ symmetry:
\bea
 \label{UNS}
X_\mu &\rightarrow & UX_\mu U^{\dagger}, \nn
\Theta& \rightarrow & U\Theta U^{\dagger},
\ena
where $U$ is an $N\times N$ unitary matrix:
\bea
UU^{\dagger} = U^{\dagger}U = \mbox{Id}_N.
\ena
The action (\ref{IIB}) also
has the following ${\cal N}=2$ 
supersymmetry:
\bea
\label{IISUSY}
\delta^{(1)} \Theta 
&=& \frac{i}{2} [X_{\mu},X_{\nu}] 
\Gamma^{\mu\nu} \epsilon^{(1)}, \nn  
\delta^{(1)} X_\mu 
&=& i\bar{\epsilon}^{(1)} \Gamma_{\mu} \Theta , \nn
\delta^{(2)} \Theta &=& \epsilon^{(2)}, \nn
\delta^{(2)} X_\mu &=& 0.
\ena
Noncommutative
Yang-Mills theory appears
when we consider
the model in a certain classical
background \cite{MLi}\cite{IKKTB}.
This background
is a  solution to the classical equation of motion,
and is identified with 
D-brane in type IIB superstring theory.
The classical equation of motion 
of the IIB matrix model is 
given by
\bea
 \label{IIBeq}
[X_{\mu},[X_{\mu},X_{\nu}]] =0.
\ena 
One class of solutions
to (\ref{IIBeq}) is given
by simultaneously diagonalizable
matrices, i.e.
$[X_{\mu},X_{\nu}] = 0$ for all $\mu,\nu$.
However
the IIB matrix model has another class of
classical solutions which are
interpreted as
D-branes in type IIB superstring theory:
\bea
 \label{half}
& &X_\mu = i \hat{\pa}_\mu \otimes \mbox{Id}_n  , \nn
& &[i\hat{\pa}_\mu, i\hat{\pa}_\nu] = - i B_{\mu\nu},
\ena
where 
$B_{\mu\nu}$ is a 
constant matrix.
$i\hat{\pa}_\mu$ is an 
infinite-dimensional matrix
because if they have only finite rank,
taking a trace of both sides of (\ref{half}) results in an
apparent contradiction.
(\ref{half}) is essentially the same as
the one appearing in (\ref{comdel}).
Therefore, we define 
``coordinate matrices" $\hat{x}^{\mu}$ from
the formula (\ref{deri}):
\bea
\hat{x}^{\mu} \equiv 
- i \theta^{\mu \nu}\hat{\pa}_{\nu},
\ena
where $\theta^{\mu\nu}$ is an inverse
matrix of $B_{\mu\nu}$.
Then, their commutation relations take the
same form as those in (\ref{noncomx}): 
\bea
 \label{xxB}
[\hat{x}^{\mu}, \hat{x}^{\nu} ] = i \theta^{\mu\nu} .
\ena
We identify these infinite-dimensional matrices with operators
acting in the Fock space $\H$.
Thus, the noncommutative coordinates of ${\R}^{2d}$
appear as a classical solution of the IIB matrix model,
where $2d$ is the rank of $B_{\mu\nu}$
and the dimension of the noncommutative
directions.
Now, let us expand the fields around
this background:
\bea
X_{\mu} &=& i (\hat{\pa}_{\mu} + A_{\mu})
        \equiv i \hat{D}_\mu \, , \label{bgf} \\
X_{I} &=& \Phi_I \, . \label{bgphi}
\ena
Here, $\mu , \nu $ are the indices of the noncommutative 
directions, i.e. $det\, \theta^{\mu\nu} \ne 0$ and
$I,J$ are the indices of the directions transverse to 
the noncommutative directions.
Then, the action (\ref{IIB}) becomes
\bea
S &=&  -\frac{1}{g^2}
\mbox{Tr}_{\H} \mbox{tr}_{U(n)} \Biggl[
- \frac{1}{4}
(F_{\mu\nu}+iB_{\mu\nu})
(F^{\mu\nu}+iB^{\mu\nu})
+
\frac{1}{2}
D_\mu \Phi_I D^{\mu} \Phi_I \Biggr. \nn
& &\qquad \qquad \qquad \quad
\Biggl. +
\frac{1}{4}
[\Phi_I, \Phi_J][\Phi_I, \Phi_J]
+
\frac{1}{2}\bar{\Theta}\Gamma^\mu
D_\mu \Theta
+
\frac{1}{2}\bar{\Theta}\Gamma^I
[\Phi_I,\Theta] \Biggr].
\ena
Here,
\bea
D_\mu \Phi_I 
&\equiv&
[\hat{D}_\mu , \Phi_I]
=
\pa_\mu \Phi_I + [A_\mu, \Phi_I ], \\
D_\mu \Theta 
&\equiv&
[\hat{D}_\mu ,\Theta]
=
\pa_\mu \Theta + [A_\mu, \Theta ] .
\ena
We thus obtain a supersymmetric noncommutative
Yang-Mills theory with a $U(n)$ gauge group. 
The gauge transformation
follows from (\ref{UNS}):
\bea
A_\mu &\rightarrow & UA_\mu U^{\dagger} 
+ U\pa_\mu U^{\dagger}, \label{GA}\\
\Phi_I &\rightarrow & U\Phi_I U^{\dagger}, \\
\Theta &\rightarrow & U\Theta U^{\dagger} .
\ena
Here, $U$ is a unitary operator:
\bea
 U U^{\dagger} = U^{\dagger} U 
= \mbox{Id}_{\H} \otimes \mbox{Id}_n .
\ena
The transformation of the gauge field (\ref{GA})
is determined by the requirement
that the form of the derivative of operators
should be kept
under the gauge transformation.

As described in the previous subsection,
we can rewrite the above matrix multiplication
using ordinary functions and the star product.

\section{Anti-Self-Dual Instantons
on Noncommutative ${\R}^4$ 
with an Anti-Self-Dual Noncommutative
Parameter}\label{ncADHM}
In this section we first review the
ADHM construction of instantons
on noncommutative ${\R}^4$, and
then use it to construct an
anti-self-dual instanton on
noncommutative ${\R}^4$ 
with an anti-self-dual noncommutative
parameter $\theta^{\mu\nu}$.
\subsection*{Review of the ADHM Construction}
The ADHM construction is a way to obtain
instanton solutions on ${\R}^4$ from solutions of
some quadratic matrix equations \cite{ADHMconst}.
It was generalized to the case of
noncommutative ${\R}^4$ in \cite{NS}.\footnote{%
For more detailed explanations on the ADHM construction
on noncommutative ${\R}^4$, see \cite{myLec}\cite{NLec}.
}
The steps in the ADHM construction
of instantons 
on noncommutative ${\R}^4$ with
noncommutative parameter $\theta^{\mu\nu}$,
gauge group $U(n)$ and instanton number $k$
is as follows:

\begin{enumerate}

\item Matrices (entries are c-numbers):  
\bea
 B_1 , B_2 &:& k \times k \quad \mbox{complex matrices.} \nn
 I, J^{\dagger} &:& k \times n \quad \mbox{complex matrices.}
\ena

\item Solve the ADHM equations:
\bea
\mu_{\R} &=& \zeta
\qquad \mbox{(real ADHM equation)},
\label{rADHM}\\ 
 \mu_{\C} &=& 0
\qquad \mbox{(complex ADHM equation)} .
\label{cADHM}
\ena
Here $\zeta \equiv 2(\theta^{12} + \theta^{34})$
and $\mu_{\R}$ and $\mu_{\C}$
are defined by
\bea
 \label{ADHMzeta}
\mu_{\R}
 &\equiv& [B_1 , B_1^{\dagger}] 
+  [B_2 , B_2^{\dagger}] 
        + II^{\dagger} - J^{\dagger} J ,\\
\mu_{\bf C} &\equiv& [B_1 , B_2 ] + IJ  .
\ena  

\item Define $2k \times (2k + n)$ matrix
${\cal D}_{z}$ :
\bea
 \label{Dz}
& &{\cal D}_{z} \equiv
\left(
 \begin{array}{c}
   \tau_{z} \\
   \sigma_{z}^{\dagger }
 \end{array}
\right) , \nn
& & \tau_{z} \equiv
(\, B_2 - {z}_2 ,\, B_1 - {z}_1 , \, I \, ), \nn
& & \sigma_{z}^{\dagger}  \equiv
( \, - (B_1^{\dagger}-\bar{{z}}_1) , \,
  B_2^{\dagger} - \bar{{z}}_2  , \, J^{\dagger} \, ) .
\ena
Here, $z$ and $\bz$ are noncommutative
{\em operators}.

\item Look for all solutions to the equation 
\bea
 \label{zeroPsi}
{\cal D}_{z} \Psi^{(a)} = 0  \quad
( a = 1, \ldots , n),
\ena
where $\Psi^{(a)} $ is a $2k+n$ dimensional vector
and its entries 
are {\em operators}. \ %
Here, we impose the following normalization condition
on $\Psi^{(a)} $:
\bea
 \label{norm}
\Psi^{(a)\dagger}\Psi^{(b)} = \delta^{ab} \mbox{Id}_{\H} \, .
\ena
In the following we will call these zero-eigenvalue vectors 
$\Psi^{(a)}$ {\bf zero-modes}.

\item Construct a gauge field by the formula
\bea
\label{ncA} 
A^{ab}_\mu
= \Psi^{(a) \dagger } \pa_\mu \Psi^{(b)} , 
\ena
where $a$ and $b$ become indices
of the $U(n)$ gauge group.
Then, this gauge field is anti-self-dual.
\end{enumerate}
From the gauge field (\ref{ncA}),
we obtain the following expression for the field
strength (for a derivation, see  for example \cite{myLec}):
\bea
\label{cFS}
& & F  \nn
&=& \!\!\!\!
\begin{array}{ccc}
\bigl(\, \psi_1^{\dagger} \,& \,
 \psi_2^{\dagger} \,& \, \xi^{\dagger} \, 
\bigr) \\
   &  &  \\
   &  &  
\end{array} \! \!
\left(
 \begin{array}{ccc}
dz_1 \frac{1}{\, \, \Box_z}d\bar{z}_1 
+ d\bar{z}_2 \frac{1}{\, \, \Box_z} dz_2   & 
  -dz_1 \frac{1}{\, \, \Box_z} d\bar{z}_2 
       + d\bar{z}_2\frac{1}{\, \, \Box_z} 
dz_1 & 0 \\
-dz_2 \frac{1}{\, \, \Box_z} d\bar{z}_1 
+ d\bar{z}_1\frac{1}{\, \, \Box_z} dz_2 & 
  dz_2 \frac{1}{\, \, \Box_z} d\bar{z}_2 
  + d\bar{z}_1 \frac{1}{\, \, \Box_z} dz_1 & 0 \\
 0 & 0 & 0
 \end{array}
\right) 
\left(
\begin{array}{c}
\psi_1 \\
\psi_2 \\
\xi
\end{array}
\right) \nn
&\equiv& F^{-}_{\mbox{\tiny ADHM}} \, ,
\ena
where we have written
\bea
& &\Psi \equiv
\left(
\begin{array}{c}
\psi_1 \\
\psi_2 \\
\xi
\end{array}
\right) \equiv
\left(
\begin{array}{ccc}
 & & \\
\Psi^{(1)} & \cdots & \Psi^{(n)} \\
 & &
\end{array}
\right) , \qquad
\begin{array}{c}
\psi_1 : k \times n \, \, \mbox{matrix.}\\
\psi_2 : k \times n \, \, \mbox{matrix.}\\
\, \,  \xi \, \,  : n \times n\, \, \mbox{matrix.}
\end{array}
\ena
In the above we have suppressed $U(n)$ gauge indices.
$F^{-}_{\mbox{\tiny ADHM}}$ is 
anti-self-dual:
$F_{1\bar{1}}+ F_{2\bar{2}} = 0$,
$F_{12} = 0$.

There is an action of $U(k)$ that does not change
the gauge field constructed by the ADHM method:
\bea
 \label{appUk}
(B_1,B_2,I,J)
\mapsto
(u B_1 u^{-1}, u B_2 u^{-1} , u I, J u^{-1}),
\qquad u \in U(k).
\ena
Therefore the moduli space ${\cal M}_{\zeta} (k,n)$ of
instantons on noncommutative 
${\R}^4$ with noncommutative parameter $\theta^{\mu\nu}$,
gauge group $U(n)$ and
instanton number $k$ is given by
\bea
  \label{moduli}
{\cal M}_{\zeta} (k,n)
 = 
\mu^{-1}_{\R}(\zeta) \cap \mu^{-1}_{\C}(0) /U(k) .
\ena
Here, the action of $U(k)$ is the one
given in (\ref{appUk}).
As stated in the previous section, in
this article we consider the case where
$\zeta = 0$. In this case the moduli
space ${\cal M}_{\zeta} (k,n)$ has so-called 
small instanton singularities which appear
when the size of the instanton becomes zero.
When $\zeta \ne 0$,
the moduli space ${\cal M}_{\zeta} (k,n)$ 
does not have  small instanton singularities \cite{iNakaj}.


In the following 
we will sometimes find it more convenient to work with 
the variable $X_\mu$ in the IIB matrix model,
rather than to work with $A_{\mu}$.
From (\ref{bgf}) and  (\ref{ncA}), we obtain the
following simple expression for the
instanton solution $X_\mu$:
\bea
X_{\mu} 
&=&  
 i\hat{\pa}_{\mu} + i A_{\mu}  = i\hat{\pa}_{\mu} 
+ i \Psi^{\dagger} \hat{\pa}_\mu \Psi
   -i \Psi^{\dagger} \Psi 
     \hat{\pa}_{\mu} \nn
&=& i \Psi^{\dagger} \hat{\pa}_\mu \Psi 
\qquad \qquad \qquad \qquad (\mu = 1,\cdots, 4).
\label{IIBD}
\ena
From (\ref{IIBD}) we obtain
\bea
 \label{XX}
[X_{\mu}, X_{\nu}]
=
- F^{-}_{\mu\nu\, \mbox{\tiny ADHM} } - i B_{\mu\nu} 
 \, ,
\ena
where
$ F^{-}_{\mu\nu\, \mbox{\tiny ADHM} } $ 
is given by (\ref{cFS}). 
From (\ref{XX}) it is easily shown that the
$X_\mu$ in (\ref{IIBD}) satisfies 
the classical equation of motion of
the IIB matrix model (\ref{IIBeq}).
It is also easy to show that
this configuration preserves one-fourth of the
supersymmetry \cite{IKKTB}.

\subsection*{$U(2)$ One-Instanton Solution and
Small Instanton Limit}

Now, let us construct an instanton by the ADHM method.
The simplest solution may be 
a $U(2)$ one-instanton solution.
In this case, $B_1$ and $B_2$ are $1\times1$ matrices,
i.e. complex numbers. Therefore, commutaters with 
$B_1$ and $B_2$ automatically give zero, and
a solution to the ADHM equation
(\ref{ADHMzeta})
is given by
\bea
 \label{1-1}
B_1 = B_2 = 0,
\quad I = (\rho \, \, 0),\quad \, J^{\dagger} = (0 \, \, \rho) .
\ena
Then, from (\ref{Dz}) we obtain
\bea
{\cal D}_z =
\left(
 \begin{array}{cccc}
  -z_2 & - z_1 & \rho & 0 \\
  \bz_1 & - \bz_2 & 0 & \rho
 \end{array}
\right) .
\ena
A solution $\Psi$ to 
the equation 
${\cal D}_z \Psi = 0 $ is given by
\bea
 \label{Psi}
& & \Psi =
\left(
 \begin{array}{cc}
        \\
  \Psi^{(1)} \, \, \Psi^{(2)} \\
 {}
\end{array}
\right), \nn
& & \Psi^{(1)}
=
\left(
 \begin{array}{c}
  \rho \\
   0 \\ 
  z_2 \\
  - \bz_1
 \end{array}
\right) \frac{1}{\sqrt{z_1\bz_1 + \bz_2 z_2+ \rho^2}}
=
\left(
 \begin{array}{c}
  \rho \\
   0 \\ 
  \sqrt{\zeta_1} a_2^{\dagger} \\
  -  \sqrt{\zeta_1} a_1^{\dagger}
 \end{array}
\right) \frac{1}{\sqrt{ \zeta_1 (\hat{N} + 2) + \rho^2}}\,
, \nn
& &\Psi^{(2)}
=
\left(
 \begin{array}{c}
  0 \\
 \rho \\ 
  z_1 \\
 \bz_2
 \end{array}
\right) 
\frac{1}{\sqrt{ \bz_1z_1 + z_2\bz_2 + \rho^2}} 
=
\left(
 \begin{array}{c}
  0 \\
 \rho \\ 
  \sqrt{\zeta_1} a_1 \\
  \sqrt{\zeta_1} a_2
 \end{array}
\right) 
\frac{1}{\sqrt{ \zeta_1 \hN + \rho^2}} \, .
\ena
Here, $\hN \equiv a_1^{\dagger} a_1 + a_2^{\dagger} a_2$.
The zero-mode $\Psi$ is normalized as in (\ref{norm}):
\bea
\Psi^{\dagger} \Psi = 
\left(
\begin{array}{cc}
 \mbox{Id}_{\H} & 0 \\ 
 0 & \mbox{Id}_{\H}
\end{array}
\right) .
\ena
The gauge field is given by (\ref{ncA}):
\bea
 \label{1A}
A_\mu (\hat{x}) = \Psi^{\dagger} \pa_\mu \Psi
\equiv A_\mu^{(0)}(\hat{x}). 
\ena
We can construct following
classical solution of the IIB matrix model:
\bea
\label{r0inst}
X_\mu &=& i \Psi^{\dagger} \hat{\pa}_\mu \Psi \qquad \quad
(\mu,\nu = 1, \cdots ,4) , \nn
X_I &=& c_I \mbox{Id}_{\H} \otimes \mbox{Id}_2 
\quad (I,J = 0, 5, \cdots 9).
\ena
This solution is interpreted as 
a system of (Euclidean) D3-brane with
NS-NS B-field background in its worldvolume and
a D(-1)-brane dissolved in the worldvolume of the
D3-branes.
From (\ref{r0inst}) we obtain
\bea
[ X_\mu , X_\nu ] &=& 
- F_{\mu\nu\, \mbox{\tiny ADHM}}^- - i B_{\mu\nu},  \nonumber \\
\, [ X_{\mu} , X_{I} ] &=& [ X_{I} , X_{J} ] = 0.
\ena
Here, $\mu, \nu$ are indices of the directions
along the worldvolume of the D3-branes
and $I,J$ are indices of the directions transverse to the
D3-branes.
The explicit form of the field strength
can be obtained from (\ref{cFS}):
\bea
 \label{FSu21}
F_{1\bar{1}\, \mbox{\tiny ADHM}}^-
= - F_{2\bar{2}\, \mbox{\tiny ADHM}}^-
&=&
\left(
 \begin{array}{cc}
 \frac{\rho^2}{(\zeta_1 (\hN + 1) + \rho^2)
(\zeta_1 (\hN + 2) + \rho^2)} & 0 \\
 0 & 
- \frac{\rho^2}{\zeta_1 (\hN + \rho^2) (\zeta_1 (\hN + 1) + \rho^2)} 
 \end{array}
\right), \nn
F_{1\bar{2}\, \mbox{\tiny ADHM}}^-
=  -F_{2\bar{1}\, \mbox{\tiny ADHM}}^{-\, \dagger}
&=&
\left(
 \begin{array}{cc}
 0 & - \frac{2\rho^2}{(\zeta_1 (\hN + 1) + \rho^2)
\sqrt{\zeta_1 (\hN + \rho^2) }  \sqrt{\zeta_1 (\hN + 2) + \rho^2} } \\
 0 & 0
 \end{array}
\right) .
\ena
From (\ref{FSu21}) one can observe
that the parameter $\rho$ characterises 
the size of the instanton.

The calculation of the instanton number
reduces to the surface integral at infinity,
and when the gauge group is $U(2)$,
the effect of the noncommutativity vanishes 
there.\footnote{
When the gauge group is $U(1)$,
the effect of the noncommutativity
do not vanish at infinity \cite{myLec}.
}
Hence the instanton number is 
classified by $\pi_3(U(2))$,
and the configuration (\ref{1A}) 
has instanton number one.
It is a little fun to study how
the direct calculation using the 
field strength (\ref{FSu21})
leads to the instanton number one.
From (\ref{FSu21}), we obtain
\bea
F_{1\bar{1}} F_{2\bar{2}}
&=&
-\frac{1}{\zeta_1}
\left(
 \begin{array}{cc}
  \frac{s^2}{(\hat{N}+1+s)^2(\hat{N}+2+s)^2} & 0 \\
    0 & \frac{s^2}{(\hat{N}+s)^2(\hat{N}+1+s)^2}
 \end{array}
\right), \nn
F_{1\bar{2}}F_{2\bar{1}}
&=&
\frac{1}{\zeta_1}
\left(
 \begin{array}{cc}
 \frac{4s^2}{(\hat{N}+s) (\hat{N}+1+s)^2 (\hat{N}+2+s)}  
       & 0 \\
 0 &  0 
 \end{array}
\right), \nn
F_{2\bar{1}} F_{1\bar{2}}
&=&
\frac{1}{\zeta_1}
\left(
 \begin{array}{cc}
 0 &  0 \\
 0 &  \frac{4 s^2}{(\hat{N}+s) (\hat{N}+1+s)^2 (\hat{N}+2+s)}
 \end{array}
\right) ,
\ena
where we have introduced a dimensionless 
parameter
$s \equiv \frac{\rho^2}{\zeta_1}$.
Then the instanton number (\ref{Inum}) becomes
\bea
& &- \frac{1}{8\pi^2}\int \mbox{tr}_{U(2)} F F \nn
&=&
(\zeta_1)^2
\mbox{Tr}_{\H} 
\mbox{tr}_{U(2)}
\left[
- F_{1\bar{1}}F_{2\bar{2}}
+ 
\frac{1}{2} 
\left(
F_{1\bar{2}}F_{2\bar{1}} +
F_{2\bar{1}}F_{1\bar{2}}
\right) \right] \nn
&=&
\mbox{Tr}_{\H} 
\frac{s^2}{(\hat{N}+1+s)^2}
\left(
\frac{1}{(\hat{N}+2+s)^2} +
\frac{1}{(\hat{N}+s)^2} +
\frac{4}{(\hat{N}+2+s)(\hat{N}+s)}
\right) \nn
&=&
\sum_{(n_1,n_2)}
\frac{s^2}{(N+1+s)^2}
\left(
\frac{1}{(N+2+s)^2} +
\frac{1}{(N+s)^2} +
2 \left( 
\frac{1}{(N+s)} - \frac{1}{(N+2+s)} 
 \right)
\right) \nn
&=&
\sum_{N=0}^{\infty}
(N+1)
\frac{s^2}{(N+1+s)^2}
\left(
 \frac{2(N+s)+1}{(N+s)^2} -
 \frac{2(N+s+1)+1}{(N+2+s)^2} 
\right) \nn
&=&
\sum_{N=0}^{\infty}
\frac{s^2}{(N+1+s)^2}
 \frac{2(N+s)+1}{(N+s)^2}\nn
& &\qquad +
\sum_{N=0}^{\infty}
s^2
\left(
N \frac{2(N+s)+1}{(N+s)^2(N+1+s)^2} 
-
(N+1) \frac{2(N+s+1)+1}{(N+1+s)^2(N+2+s)^2} 
\right)  \nn
&=&
\sum_{N=0}^{\infty}
\frac{s^2}{(N+1+s)^2}
 \frac{2(N+s)+1}{(N+s)^2}  \nn
&=&
\sum_{N=0}^{\infty} \, \, 
\frac{s^2}{(N+s)^2} - \frac{s^2}{(N+1+s)^2} 
=
1.
\ena
Thus the instanton number is one, 
independent of the parameter $\rho$ 
which characterizes the size of the instanton.

Now, let us consider the small instanton limit, i.e.
$\rho \rightarrow 0$.
The moduli space (\ref{moduli}) becomes singular
at $\rho = 0$.
When $\rho = 0$, the zero-mode $\Psi$ takes the 
following form:
\bea
 \label{Psir0}
& &\Psi =
\left(
 \begin{array}{cc}
        \\
  \Psi^{(1)} \, \, \Psi^{(2)} \\
 {}
\end{array}
\right), \nn
\Psi^{(1)}
&=&
\left(
 \begin{array}{c}
   0 \\
   0 \\ 
  a^{\dagger}_2 \\
  - a^{\dagger}_1
 \end{array}
\right) \frac{1}{\sqrt{ \hN +2 }}
\, , \quad
\Psi^{(2)}
=
\left(
 \begin{array}{c}
  0 \\
\left| 0,0 \right\ran \left\lan 0,0\right| \\ 
  a_1 \frac{1}{\sqrt{ \hN_{\ne 0} } } \\
  a_2 \frac{1}{\sqrt{ \hN_{\ne 0} } }
 \end{array}
\right) \, ,
\ena
where $\frac{1}{\sqrt{ \hN_{\ne 0} } }$ is defined as
\bea
\frac{1}{\sqrt{ \hN_{\ne 0} } } \equiv 
\sum_{(n_1,n_2)\ne (0,0)}
\frac{1}{\sqrt{n_1+n_2}} 
\left| n_1,n_2 \right\ran \left\lan n_1,n_2\right|\, .
\ena
Thus when $\rho = 0$, the explicit form of 
the gauge field is given by
\bea
 \label{A0}
A_{\mu} (\hat{x}) = 
U^{\dagger} \pa_\mu U + 
\left| 0,0 \right\ran \left\lan 0,0\right|
\pa_\mu
\left| 0,0 \right\ran \left\lan 0,0\right|,
\ena
where 
\bea
 \label{PI}
U 
\equiv
\frac{1}{\sqrt{\hN + 1}}
\left(
 \begin{array}{cc}
 a_2^{\dagger} & a_1 \\
 -a_1^{\dagger} &   a_2
 \end{array}
\right) .
\ena
$U$ satisfies the following equations:
\bea
 U U^{\dagger} = \mbox{Id}_{\H}\otimes \mbox{Id}_2,\quad
 U^{\dagger} U = 
\left(
 \begin{array}{cc}
  \mbox{Id}_{\H} & 0 \\ 
  0 & \mbox{Id}_{\H} - 
 \left| 0,0 \right\ran \left\lan 0,0\right|
 \end{array}
\right) \equiv p.
\ena
Note that $p$ is a projection operator:
$p^2 = p, \, p^{\dagger} = p $.
The field strength becomes
\bea
 \label{sFS}
F_{\mu\nu} (\hat{x}) = i (1-p) B_{\mu\nu}  
=
\left(
 \begin{array}{cc}
  0 & 0 \\ 
  0 &  \left| 0,0 \right\ran \left\lan 0,0\right|
 \end{array}
\right) i B_{\mu\nu}  
\equiv F_{\mu\nu}^{(0)} (\hat{x}).
\ena
Note that
$\Omega_W (\left| 0,0 \right\ran \left\lan 0,0\right|)
 = 4 e^{-\frac{2}{\zeta_1} r^2}
$, 
where $r^2 \equiv x_\mu x^\mu$.
Thus the $\rho = 0$
corresponds to the ``minimal size'' 
\footnote{The functional form of the
gauge field depends on gauge choice.}
instanton.
The Weyl symbol of the
field strength in this case
is a
Gaussian function concentrated at the origin,
with spreading of order $\sim \sqrt{\zeta_1}$.
It is explicitly non-singular.

In terms of the IIB matrix model variable,
we obtain
\bea
 \label{agms}
X_\mu =  i U^{\dagger} \hat{\pa}_\mu U 
\equiv X_\mu^{(0)}.
\ena
In the above we have used the equation
$\left\lan 0,0\right|
\hat{\pa}_\mu
\left| 0,0 \right\ran = 0$.

Since $U = Up$ and $U^{\dagger} = p U^{\dagger}$,
one can add new parameters to the 
$\rho = 0$ solution (\ref{agms}) without
changing the field strength (\ref{sFS}) 
:
\bea
 \label{Clo}
X_\mu &=& i U^{\dagger} \hat{\pa}_\mu U + 
         c_\mu ( 1 - p), \nn
X_I &=& C_I p + c_I ( 1 - p) ,
\ena
where we have introduced 
c-number parameters
$c_\mu$ and $c_I$.
$c_{\mu}$'s are related to 
the parameters of the ADHM moduli space,
as we will explain in the next subsection.
The field strength is unchanged
by the modification from (\ref{r0inst}) to (\ref{Clo}), 
in particular, it remains anti-self-dual.
Hence the configuration (\ref{Clo}) 
is still a solution of the IIB matrix model and
preserves one-fourth of the supersymmetry.
Since the projections 
$p$ and $1-p$ are orthogonal,
we can express the solution 
(\ref{Clo}) in a block diagonal form.
Taking an appropriate basis, we can write,
schematically,
\bea
 \label{block}
X_\mu &=&
 \left(
  \begin{array}{cc}
   i \hat{\pa}_\mu \otimes \mbox{Id}_2 & {} \\
   {} & c_\mu 
  \end{array}
 \right) , \nn
X_I &=& 
\left(
  \begin{array}{cc}
   C_I \mbox{Id}_{\H} \otimes \mbox{Id}_2 & {} \\
   {} & c_I 
  \end{array}
 \right).
\ena
Thus, the solution (\ref{Clo}) is
the same as that discussed in \cite{IKKT}\cite{IIBD}
and recently considered in \cite{AGMS} in the context of
the noncommutative Yang-Mills theory.
In the IIB matrix model,
the left-upper block in (\ref{block}) is interpreted
as D3-branes, and the right-lower block is interpreted
as a D(-1)-brane.
The parameter set $(c_\mu, c_I)$ is interpreted
as a position of the D(-1)-brane,
and the parameter $C_I$ is interpreted
as a position of the D3-branes.
When $C_I - c_I \ne 0$, the solution (\ref{Clo})
describes the D(-1)-brane and the D3-branes
separated in the direction transverse to the
D3-branes.

\subsection*{Comments on the ``Position'' of the Instanton}

There is an obvious extension of the solution
(\ref{1-1}) which follows from the translational
symmetry on noncommutative ${\R}^4$:
\bea
 \label{w1-1}
B_1 = w_1,\, \, B_2 = w_2, 
\quad I = (\rho \, \, 0),\quad \, J^{\dagger} = (0 \, \, \rho).
\ena
Then, from (\ref{Dz}) we obtain
\bea
 \label{tDz}
{\cal D}_z =
\left(
 \begin{array}{cccc}
  -(z_2-w_2) & - (z_1-w_1) & \rho & 0 \\
  (\bz_1- \bw_1) & - (\bz_2 - \bw_2) & 0 & \rho
 \end{array}
\right) .
\ena
Following the steps 
parallel to the previous subsection,
we obtain a gauge field 
which is obtained from 
$A_{\mu}^{(0)} (\hat{x})$ in 
(\ref{1A}) by translation:
\bea
 \label{traA1}
A_{\mu}(\hat{x}) = 
A_{\mu}^{(0)}(\hat{x} - y).
\ena
Here, $y$'s are defined from 
\bea
w_1 \equiv y^2 + i y^1, \quad w_2 \equiv  y^4 + i y^3 .
\ena
The field strength becomes
\bea
 \label{Fy}
F_{\mu\nu}(\hat{x}) = F^{(0)}_{\mu\nu}(\hat{x} - y),
\ena
where $F^{(0)}_{\mu\nu}(\hat{x})$
is the one defined in (\ref{sFS}).
Thus the parameter $y^{\mu}$ can be interpreted as 
position of the instanton on noncommutative ${\R}^4$.
This is parallel to the interpretation 
in the commutative case. 
However, on noncommutative ${\R}^4$,
the notion of the position 
should be considered with care \cite{GHI}, 
since the 
translation (\ref{tra})
generates unitary
gauge transformation on
noncommutative ${\R}^4$: 
\bea
 \label{traA2}
A_{\mu}'(\hat{x})
= T_{y} A_{\mu}^{(0)}(\hat{x}-y) T^{\dagger}_{y} +
  T_{y} \pa_{\mu} T^{\dagger}_{y}
= A_{\mu}^{(0)}(\hat{x}) - i B_{\mu\nu} y^{\nu} .
\ena
Thus the difference between $A_{\mu}'(\hat{x})$,
the gauge transform of $A_{\mu}$ in (\ref{traA1}),
 and
$A_{\mu}^{(0)}(\hat{x})$ is constant.
Both  
the translation and
the constant shift of the gauge field
are symmetries 
of the action (\ref{SO}) and 
(\ref{IIB}).
The eq. (\ref{traA2})
means that these two symmetries of the
action do not
lead to independent moduli parameters,
but they are related by the gauge transformation.


Let us study the solution in the 
framework of the IIB matrix model.
From the ADHM data (\ref{w1-1}), we 
obtain a classical solution of the 
IIB matrix model:
\bea
 \label{XmuT}
X_\mu(\hat{x}) = X_{\mu}^{(0)}(\hat{x} - y) + c_\mu ,
\ena
where
\bea
 \label{target}
c_\mu \equiv - B_{\mu\nu} y^{\nu}
\qquad (\mu, \nu = 1, \cdots , 4),
\ena
and $X_{\mu}^{(0)}(\hat{x})$ is defined in (\ref{agms}).
Using an appropriate basis in $\H$, $X_\mu(\hat{x})$ 
can be written in a block diagonal form just like 
$X_{\mu}^{(0)}(\hat{x})$ in (\ref{agms}):
\bea 
 \label{block2}
X_\mu (\hat{x}) &=&
 \left(
  \begin{array}{cc}
   (i \hat{\pa}_\mu - c_\mu) \otimes \mbox{Id}_2 & {} \\
   {} & c_\mu 
  \end{array}
 \right) , \nn
X_I &=& 
\left(
  \begin{array}{cc}
   C_I \mbox{Id}_{\H} \otimes \mbox{Id}_2 & {} \\
   {} & c_I 
  \end{array}
 \right).
\ena
The configuration (\ref{XmuT}) is gauge equivalent to (\ref{Clo}).
To see this, let us first consider
the unitary transformation $T_{y}$. 
By this unitary transformation, we obtain a
gauge equivalent configuration $X_\mu'$:
\bea
X_{\mu}'(\hat{x})
= T_{y} \left( X_{\mu}^{(0)}(\hat{x} - y)+ c_\mu \right) 
T^{\dagger}_{y}  
= X_{\mu}^{(0)}(\hat{x}) + c_\mu .
\ena
Next, let us consider the following  unitary operator $V$:
\bea
V \equiv \left( U^{\dagger}T_{-y} U + (1-p) \right)  T_{y} .
\ena
By this unitary transformation $V$, we obtain $X_\mu''$
which is gauge equivalent to $X_\mu$ in (\ref{XmuT})
with $\rho = 0$:
\bea
 \label{IIBgauge}
X_\mu''(\hat{x})=
V \bigl. X_{\mu}  (\hat{x}) \bigr|_{\rho = 0} V^{\dagger} 
=
i U^{\dagger} \hat{\pa}_\mu U 
+ c_\mu (1-p) .
\ena
This is the expression appeared in (\ref{Clo}),
and there $c_\mu$ is interpreted as a position of the
D(-1)-brane in the worldvolume of the D3-branes.
The equation (\ref{target}) means that
the ``position'' parameter $y^{\mu}$
in the ADHM moduli
is essentially equivalent to the 
 ``position'' $c_\mu$
of the D(-1)-brane, in the direction parallel
to the worldvolume of the D3-branes.\footnote{%
If we set 
$X^\mu = \hat{x}^\mu + \theta^{\mu\nu} A_\nu$
instead of (\ref{bgf}), then
the position parameter in the ADHM moduli 
exactly coincides with 
the position of the D(-1)-brane.
Therefore, the difference in $y$ and $c$
in (\ref{target}) is only a matter of 
the choice of the coordinate system. 
}

These position parameters will enter 
in the gauge invariant
observables like the ones considered
in \cite{GHI}\cite{Dhar}.

\subsection*{Multi-Small Instanton Solution}

In general, it is quite difficult to obtain
an explicit expression of the zero-modes (\ref{zeroPsi})
in the noncommutative version of the
ADHM construction.
The reason is as follows.
When all the instantons are top on each other
at the origin, then we can 
use the basis (\ref{basis}) of the
Fock space $\H$.
\footnote{
In \cite{NLec}
the explicit multi-instanton solutions 
with $U(1)$ gauge group is obtained.
This is because 
in the case of the anti-self-dual instantons on
self-dual noncommutativity,
we can utilize 
a noncommutative analogue of ``singular gauge'',
which simplifies the problem of finding zero-modes.
But such a gauge choice 
cannot be used here. The reason may be 
understood from the explanation in \cite{myLec},
sec.4.3.
}
However, when the instantons are ``separated'',
there is no such convenient basis
(
the meaning of the word ``separated''
here will be made clearer shortly).
However, when all  the instantons
become small instantons, i.e. $I = J^{\dagger} = 0$,
we can construct an explicit $k$-instanton solution
quite easily:
\bea
 \label{multi0}
B_1 =
\left(
 \begin{array}{cccc}
  w_1^{(1)} & 0 &\cdots &  0 \\
  0 &  w_1^{(2)} &  & \vdots \\
  \vdots &  & \ddots & 0 \\
  0 &  \cdots & 0 & w_1^{(k)}
 \end{array}
\right) , \quad 
B_2 =
\left(
 \begin{array}{cccc}
  w_2^{(1)} & 0 &\cdots &  0 \\
  0 &  w_2^{(2)} &  & \vdots \\
  \vdots &  & \ddots  & 0 \\
  0 &  \cdots & 0 & w_2^{(k)}
 \end{array}
\right) , \nonumber
\ena
\bea
I = J^{\dagger} = 0 .
\ena
%
Here $w^{(i)}$ is a 
parameter that expresses the
position of the $i$-th instanton.
When $w^{(i)} \ne w^{(j)}$, we will state that
the $i$-th instanton and $j$-th 
instanton are separated.

We can construct a zero-mode corresponding to
(\ref{multi0}):
\bea
 \label{mPsir0}
&\overbrace{\qquad \qquad \qquad \quad  }^{\textstyle 2}& \nn
&\Psi =
\left(
 \begin{array}{l}
  0   \qquad \qquad  \qquad 0 \\
 \, \vdots \qquad \qquad  \qquad  \vdots \\
  0 \qquad \qquad  \qquad 0 \\
  0   \qquad  | w_1^{(1)}, w_2^{(1)}\ran \left\lan 0,0\right| \\
 \, \vdots \qquad \qquad  \qquad  \vdots \\
 0  \quad     | w_1^{(k)}, w_2^{(k)}\ran \left\lan 0,k-1\right| \\
 \ \ \ \ \  \ \ \ \ \Biggl. U^{k} \Biggr.
\end{array}
\right) 
& 
\begin{array}{l}
 {} \\
 \Biggl. {} \Biggr\} k \\
 {} \\
 \Biggl. {} \Biggr\} k \\
 {} \\
 \}2  \, \, . 
\end{array} \\
{}
\ena
Here, the numbers above and 
the right hand side of the 
matrix denote the number of the columns
and the number of the lines, respectively.
$| w_1^{(i)}, w_2^{(i)}\ran$ is a coherent state:
\bea
& &z_1 \, | w_1^{(i)}, w_2^{(i)}\ran =
w_1 ^{(i)}\, | w_1^{(i)}, w_2^{(i)}\ran , \nn
& &\bz_2 \, | w_1^{(i)}, w_2^{(i)}\ran =
\bw_2^{(i)} \, | w_1^{(i)}, w_2^{(i)}\ran , \nn
& &  \lan w_1^{(i)}, w_2^{(i)} | w_1^{(i)}, w_2^{(i)}\ran = 1 .
\ena
In the above, we have already
chosen a gauge similar to the one in (\ref{IIBgauge}).
To recognize that the zero-mode (\ref{mPsir0})
is the correct one, 
we recommend the reader to check the
equations
$\Psi^{\dagger}\Psi = \mbox{Id}_{\H} \otimes \mbox{Id}_2$
and
$\Psi\Psi^{\dagger} = 
1 - 
{\scriptstyle \cal D}_z^{\dagger} 
\frac{1}{ {\cal D}_z {\cal D}_z^{\dagger} }  
{\scriptstyle \cal D}_z $,
which are necessary conditions in the
ADHM construction (see for example, \cite{myLec}).
To check these equations, one can utilize 
the following equations:
\bea
U^{k}  (U^{\dagger})^{k} = \mbox{Id}_{\H} \otimes \mbox{Id}_2, \quad
(U^{\dagger})^{k}U^{k} =
\left(
 \begin{array}{cc}
  \mbox{Id}_{\H} & 0 \\ 
  0 & \mbox{Id}_{\H} - \sum_{n_2 = 0}^{k-1}
 \left| 0, n_2 \right\ran \left\lan 0,n_2\right|
 \end{array}
\right).
\ena
The IIB matrix variable $X_\mu$
takes the block diagonal form:
\bea
 \label{mblock}
X_\mu &=&
 \left(
  \begin{array}{cc}
   i \hat{\pa}_\mu \otimes \mbox{Id}_2 & {} \\
   {} & c_\mu 
  \end{array}
 \right) , \nn
X_I &=& 
\left(
  \begin{array}{cc}
   C_I \mbox{Id}_{\H} \otimes \mbox{Id}_2 & {} \\
   {} & c_I 
  \end{array}
 \right),
\ena
where
\bea
c_{\mu} = 
 \left(
 \begin{array}{cccc}
   c_{\mu}^{(1)} & 0 &\cdots &  0 \\
  0 & c_{\mu}^{(2)} &  & \vdots \\
  \vdots &  & \ddots & 0 \\
  0 &  \cdots & 0 & c_{\mu}^{(k)}
 \end{array}
\right) ,
\ena
\bea
& &c_{\mu}^{(i)} \equiv - B_{\mu\nu} y^{\nu\, (i)}, \nn
& &w_1^{(i)}  \equiv y^{2\, (i)} + i y^{1\, (i)}, 
\quad w_2^{(i)}  \equiv  y^{4\, (i)} + i y^{3\, (i)} ,
\ena
and
$c_I$'s are $k \times k$ matrices which
commute with $c_\mu$ and among themselves.
In the IIB matrix model,
the $k \times k$ blocks $c_\mu$ and $c_I$ are
interpreted as
the worldvolume of $k$ D(-1)-branes.
When
$c^{(1)} = c^{(2)} = \cdots = c^{(k)}$,
$U(k)$ symmetry enhancement occurs.
Note that this $U(k)$ symmetry is the unbroken subgroup of
the $U(\infty)$ unitary group acting in the Fock space $\H$.
This symmetry enhancement 
is parallel to the 
symmetry enhancement 
in the solution of the 
ADHM equation (\ref{multi0}), which is the
origin of the 
singularities in the moduli space.
It is interesting that
in the noncommutative case,
the origin of the singularities in the moduli space
can be 
directly observed from the field configuration.
This is owing to the fact that
the field configuration is non-singular
in the noncommutative case,
as opposed to the commutative case.

\section{Discussions}

In this article we have 
constructed an anti-self-dual instanton solution
on noncommutative ${\R}^4$ with
an anti-self-dual noncommutative parameter $\theta^{\mu\nu}$.
The solution is constructed by the ADHM construction,
and it is discussed
in the framework of the IIB matrix model.
The solution has a parameter $\rho$ that characterizes the
size of the instanton.
The case $\rho = 0$ corresponds to the small instanton 
singularity in the moduli space.
It is
shown that even at this small instanton singularity,
the solution itself is explicitly non-singular,
and takes the special form (\ref{agms}).
Then the solution is continuously
connected to 
the solution (\ref{Clo}),
which is interpreted as a
system of separated
Dp-brane and D(p+4)-branes.
This is consistent with 
an analysis of the moduli space of field theory on the
worldvolume of the Dp-brane, since in this case
the Higgs branch and the Coulomb branch are connected
at the small instanton singularity.
It is quite remarkable that while
instantons in ordinary Yang-Mills theory only
describe Dp-branes within the worldvolume
of D(p+4)-branes,
the noncommutative Yang-Mills theory can
describe the separation of 
Dp-branes off D(p+4)-branes.

We also observed that 
the instanton position parameter
essentially coincides
with the position of the 
D(-1)-brane in the IIB matrix model.
It may be interesting to investigate
the appearance of the instanton 
position parameter
in the dual supergravity side 
\cite{AdSI},
in the large $N$ super Yang-Mills/supergravity
correspondence
\cite{AdS}.

A class of multi-instanton solutions
is also constructed.
It is shown that
he origin of 
symmetry enhancement in the worldvolume theory
of D(-1)-branes 
can be observed directly
from the symmetry enhancement 
in the IIB matrix model  variable.

It will be interesting  
to clarify the
precise relation to the sigma model 
analysis, like those in \cite{SW}\cite{AGMS}.

\vs{5}
\begin{center}
{\large \bf Acknowledgements} 
\end{center}
I would like to thank 
K. Okuyama and Y. Kimura 
for explanations,
and N. Ishibashi for
inspiring questions and encouragements.
I would also like to thank 
H. Aoki, T. Hirayama, 
Y. Okada, T. Tada and
especially S. Iso
for discussions while I was 
preparing the revised version 
of this article. 
I am also benefited
from conversations with 
M. Hamanaka, 
R. Gopakumar and S. Minwalla
in the 5th Winter School of APCTP/KIAS 
and 9th Haengdang Workshop on 
Strings and Field Theory.

\newpage
\appendix

\section{%
Smoothness of the Weyl Symbol of the 
Instanton Gauge Field}


The components of the 
instanton configuration 
(\ref{1A})
is well defined as an operator
acting in the Fock space $\H$,
and free from divergences.
However, one may wonder
whether the configuration is
non-singular as a function on ${\R}^4$
after the Weyl map (\ref{WeylT}).
In this appendix, we explain how to show
the smoothness of the 
Weyl symbol of the 
instanton gauge field (\ref{1A}).
Hereafter we set $\zeta_1 = 1$ for simplicity,
but the conclusion is the same for general 
$\zeta_1$.

Let us first recall
that
the Weyl symbol of the 
projection operator 
$\left|n_1, n_2  \right\ran
\left\lan n_1, n_2 \right|$ 
is given by \cite{GMS}
\bea
 \label{Weylmn}
\Omega_W(
\left|n_1, n_2  \right\ran
\left\lan n_1, n_2 \right| )
=
2^2(-1)^{n_1 + n_2} 
L_{n_1} (4 r_1^2) L_{n_2} (4 r_2^2) e^{- 2r^2} ,
\ena
where
\bea
r_{\ }   \equiv \sqrt{(r_1)^2 + (r_2)^2}, 
& & r_1 \equiv \sqrt{(x^1)^2 + (x^2)^2 }, \nn
& & r_2 \equiv \sqrt{(x^3)^2 + (x^4)^2 },
\ena
and $L_n(x)$ is the $n$-th
Laguerre polynomial:
\bea
L_n (x) \equiv 
\frac{e^x}{n!} \frac{d^n}{ dx^n} x^n e^{-x}   .
\ena
Since for finite $n_1$ and $n_2$
the Weyl symbol (\ref{Weylmn})
is finite and infinite times differentiable,
potential dangerous of
singularity only comes from the
infinite summation over $n_1$ or $n_2$.
This means that for the discussion on
the smoothness
of the Weyl symbols,
we only need to study
the region
$n_1 + n_2 \equiv N \geq N_c$
for some large fixed integer $N_c$.

Let us write $A_\mu = A_\mu^a t^a$,
where $t^a$'s are generators of the 
$U(2)$ gauge group.
From direct calculation,
it can be shown that
the gauge configuration (\ref{1A})
can be written in the form
\bea
 \label{geneA}
A_\mu^a 
=
\hat{f}^{(1)} \hat{P}^{(1)}
+
\hat{f}^{(2)} \hat{P}^{(3)},
\ena
where $\hat{P}^{(1)}$ and $\hat{P}^{(3)}$ are
polynomials in $\hat{x}$, and 
of order equal or less than 
one and three, respectively.
Furthermore, $\hat{f}^{(1)}$ and $\hat{f}^{(2)}$
can be written as 
\bea
\hat{f}^{(1)}
&=&
\sum_{(n_1,n_2)} f_{n_1 n_2}^{(1)}
\left|n_1, n_2  \right\ran
\left\lan n_1, n_2 \right| ,\nn
\hat{f}^{(2)}
&=&
\sum_{(n_1,n_2)} f_{n_1 n_2}^{(2)}
\left|n_1, n_2  \right\ran
\left\lan n_1, n_2 \right| ,
\ena
and have the asymptotic property 
\bea
 f_{n_1 n_2}^{(1)} & \rightarrow & O(\frac{1}{N}) 
    \quad (N \rightarrow \infty) , \nn
 f_{n_1 n_2}^{(2)} & \rightarrow & O(\frac{1}{N^2}) 
    \quad (N \rightarrow \infty) .
\ena
From the explicit form (\ref{geneA}),
we observe that
to show the smoothness of the
Weyl symbol of the 
gauge configuration
(\ref{geneA}),
we only need to show
that
$\hat{f}^{(1)}$ is
two times differentiable
and
$\hat{f}^{(2)}$
is four times differentiable.
Let us write
\bea
\Box^l \hat{f}^{(s)}
= 
\sum_{(n_1,n_2)} 
(\Box^l f^{(s)})_{n_1 n_2}
\left|n_1, n_2  \right\ran
\left\lan n_1, n_2 \right| \quad 
(s = 1,2),
\ena
where $\Box \equiv \pa_\mu \pa^\mu$.
Then we can check that
\bea
 \label{derideri}
(\Box^l f^{(s)})_{n_1n_2}
\rightarrow O(\frac{1}{N^{(l+s)} }),
\ena
for arbitrary non-negative integer $l$.
From (\ref{derideri}) we obtain
\bea
 \label{bounded}
\left|
\Omega_W (
\sum_{n_1+ n_2 \geq N_c} 
(\Box^l f^{(s)})_{n_1n_2}
\left|n_1, n_2  \right\ran
\left\lan n_1, n_2 \right|
)\right|
&<&
C \sum_{n_1+ n_2 \geq N_c} 
\frac{1}{N^{(l+s)}}
\left|
\Omega_W
(\left|n_1, n_2  \right\ran
\left\lan n_1, n_2 \right|) 
 \right|\nn
&\leq& 
C
\sum_{n_1+ n_2 \geq N_c} 
\frac{1}{N^{(l+s)}}  \nn
&<&
C' \qquad (l+s > 2),
\ena
for some constants $C$ and $C'$.
In the above we have used the
inequality
$| L_n(x) e^{-x/2}| \leq 1$.
(\ref{bounded}) means that
$\hat{f}^{(s)}$ is $2 l$ times differentiable.
Since $l$ is an arbitrary non-negative integer,
this means 
$\hat{f}^{(s)}$ is infinite times differentiable.
Thus the Weyl symbol of the gauge 
configuration (\ref{geneA})
is smooth.

Thus we have shown
the smoothness of the gauge field (\ref{1A}). 
Note that
even the configuration
that corresponds to the singularity of 
the moduli space is smooth.
Since even the smallest size instanton 
is smooth in the above case,
the smoothness
of the Weyl symbols
of more general configurations
can be expected to be shown in a similar manner,
though it needs more 
precise arguments
since we cannot write the configurations
in the form (\ref{geneA})
in general.
We left the complete proof 
of the smoothness of the
Weyl symbols of more general 
instanton configurations
to the future.


\newpage
\newcommand{\NP}[1]{Nucl.\ Phys.\ {\bf #1}}
\newcommand{\AP}[1]{Ann.\ Phys.\ {\bf #1}}
\newcommand{\PL}[1]{Phys.\ Lett.\ {\bf #1}}
\newcommand{\CQG}[1]{Class. Quant. Gravity {\bf #1}}
\newcommand{\CMP}[1]{Commun.\ Math.\ Phys.\ {\bf #1}}
\newcommand{\PR}[1]{Phys.\ Rev.\ {\bf #1}}
\newcommand{\PRL}[1]{Phys.\ Rev.\ Lett.\ {\bf #1}}
\newcommand{\PRE}[1]{Phys.\ Rep.\ {\bf #1}}
\newcommand{\PTP}[1]{Prog.\ Theor.\ Phys.\ {\bf #1}}
\newcommand{\PTPS}[1]{Prog.\ Theor.\ Phys.\ Suppl.\ {\bf #1}}
\newcommand{\MPL}[1]{Mod.\ Phys.\ Lett.\ {\bf #1}}
\newcommand{\IJMP}[1]{Int.\ Jour.\ Mod.\ Phys.\ {\bf #1}}
\newcommand{\JHEP}[1]{JHEP\ {\bf #1}}
\newcommand{\JP}[1]{Jour.\ Phys.\ {\bf #1}}

\end{document}